# First-Principles Identification of Higher-Order Configurational Energy Terms and Their Magnetic-State Dependence in Lithium-Rich Oxyfluorides: From QUBO to HUBO

**Joonho Bae**[1,2,*]

[1] Department of Physics and Semiconductor Science, Gachon University, Seongnam 13120, Republic of Korea

[2] Department of Semiconductor Engineering, Gachon University, Seongnam 13120, Republic of Korea

* Corresponding author: baejh2k@gachon.ac.kr

## Abstract

In this work, we use spin-polarized density functional theory (DFT) to identify substantial third-order, nonadditive configurational energy contributions in lithium-rich oxyfluorides that quadratic unconstrained binary optimization (QUBO) models cannot represent. We test the extension from QUBO to higher-order unconstrained binary optimization (HUBO) in two disordered-rocksalt cathode materials, $Li_2MnO_2F$ (LMOF) and $Li_2Mn_{0.5}Ti_{0.5}O_2F$ (LMTiOF), using composition-preserving DFT energy maps. Using fixed-geometry energy maps, we show that cubic HUBO captures nonadditive contributions absent from QUBO models and substantially reduces fitting errors in both materials. Direct DFT energy differences confirm third-order nonadditivity, while comparisons of magnetic initializations reveal a strong dependence of the LMTiOF contribution on the converged electronic state. These results support the inclusion of higher-order terms when modeling configurational energetics and highlight the need to account for magnetic-state dependence in constructing DFT energy maps. The extracted terms describe effective couplings among composition-preserving exchange variables, rather than uniquely identified microscopic three-atom interactions.

Keywords: density functional theory; disordered rocksalt oxyfluorides; configurational energetics; higher-order nonadditivity; magnetic initialization; cluster expansion

## 1 Introduction

Lithium-rich disordered rocksalt oxyfluorides offer a chemically flexible setting for investigating how local atomic arrangements influence the energies of battery materials. In these compounds, an average rocksalt description permits many distributions of Li and transition metals over the cation sublattice, together with different O/F arrangements on the anion sublattice. Equal overall compositions therefore need not imply equivalent local environments or equal energies. The relevant first-principles question is how the energetic cost of one occupation change depends on the other occupation changes already present in the cell. Resolving this dependence is a necessary step toward constructing useful statistical descriptions of configurational disorder.

For an occupation-based model, the central issue is whether a local rearrangement has a nearly environment-independent cost or whether its cost changes when neighboring chemical assignments change. In a Mn/Ti oxyfluoride, a cation exchange and an O/F exchange alter different parts of the same self-consistent electronic problem. Their combined response is therefore a useful test of an assumed quadratic description. We use disordered rocksalt oxyfluorides as the materials context while restricting every numerical conclusion to the explicitly enumerated cells.

Mn-containing lithium-excess oxyfluorides are particularly interesting because their redox chemistry can involve substantial changes in the Mn electronic state. Lee and co-workers demonstrated reversible $Mn^{2+}/Mn^{4+}$ double redox in lithium-excess cathode materials, including the $Li_2Mn_{0.5}Ti_{0.5}O_2F$ family [1].

These findings motivate our investigation of how atomic arrangements affect configurational energies at fixed composition.

Existing DFT studies of disordered rocksalt materials provide a more direct conceptual context than optimization-algorithm benchmarks. McColl and co-workers examined $Li_2MnO_2F$ in a study of transition-metal migration, molecular oxygen formation, and voltage hysteresis [2]. Li and co-workers combined atomic-scale characterization with cluster-expansion Monte Carlo modeling to investigate short-range order in cation-disordered cathodes [3]. Squires and Scanlon used cluster-expansion-driven simulations to assess limits to suppressing short-range order through multicomponent substitution [4]. These studies demonstrate the value of connecting first-principles energies to physically defined arrangements.

It is also important to distinguish changes in which atoms occupy each site from changes in atomic positions. Szymanski and co-workers showed, through pair distribution function analysis and first-principles structural modeling, that configurational environments and local bond-length relaxation can act together in disordered rocksalt oxyfluorides [5]. A fixed-coordinate energy map isolates one controlled geometrical slice of this problem. It does not provide the relaxed energies of all corresponding arrangements. This restricted setting nevertheless permits a transparent examination of electronic nonadditivity, provided that structural strain and the missing relaxation response are treated as limitations of the physical inference.

Cluster expansion supplies the established framework for representing an energy function of discrete occupations [6–8]. Pair, triplet, and higher-cluster terms are not new ingredients in this framework. Barroso-Luque and Ceder [9] developed a basis-independent cluster decomposition of configurational energies, clarifying the distinction between expansion coefficients and physically interpretable multi-body contributions. Here, the variables are composition-preserving exchanges rather than individual lattice occupations. A polynomial in these variables provides a compact and fully specified representation on selected finite domains, but a term involving three exchange variables is not automatically a spatial triplet interaction among three atoms. The distinction between mathematical polynomial degree and microscopic interaction order is central to interpreting the calculations.

Magnetic electronic structure adds another level of dependence. An SCF calculation at fixed atomic coordinates may converge to different local solutions when the starting spin polarization changes. Explicit treatments of coupled magnetic and chemical degrees of freedom have been developed both in magnetic cluster expansion [10] and, more recently, for the spinel cathode $LiNi_{0.5}Mn_{1.5}O_4$ [11]. The latter study concerns a different crystal structure and constructs a statistical magnetic description; it is methodological context rather than direct evidence for the present oxyfluorides. Our calculations instead test how changing the initial Mn polarizations alters the energy labels assigned to the same ionic configurations.

The five materials studies cited above define complementary questions for the present work. McColl et al. [2] connect rearrangements to oxygen-redox processes, Li et al. [3] and Squires and Scanlon [4] address configurational statistics and short-range order, and Szymanski et al. [5] connect occupations with local structural response. García Ponte et al. [11] show how magnetic degrees of freedom can enter a first-principles statistical treatment in a different cathode structure. Here we examine an earlier modeling step: whether the energies supplied to an occupation-only description have a consistent electronic-state definition.

Recent configuration-search studies connect this representation question to optimization algorithms. Couzinié et al. incorporated n-body atomic interactions into annealing-based crystal-structure search [12]. Ichikawa et al. used a first-principles cluster-expansion model in numerical QAOA studies of Au–Cu configurations [13]. Their more recent work treats composition-constrained cubic and quartic cluster-expansion objectives directly with simulated bifurcation [14]. The latter is a classical optimization study, not a quantum-hardware demonstration. These studies establish prior use of higher-order configurational

Hamiltonians; the present contribution is the direct DFT diagnosis of nonquadratic response on specified oxyfluoride domains and its dependence on electronic initialization.

Higher-order objectives are also the subject of dedicated quantum-algorithm studies. Pelofske and co-workers compared QAOA with quantum annealing on higher-order Ising instances [15] and examined QAOA scaling and parameter transfer on heavy-hex graphs [16]. Romero, Hegade and co-workers developed bias-field digitized counterdiabatic optimization for HUBO and tested three-local Ising objectives on quantum hardware [17]. Koch et al. studied resource savings from higher-order encodings in gate-based optimization [18], Nagies, Hauke and co-workers compared direct polynomial and quadratized annealing formulations [19], and Kiktenko et al. investigated Grover-mixer QAOA for HUBO numerically [20].

We examine the progression from a quadratic QUBO energy representation to a cubic HUBO representation [21] using DFT configurational maps, direct higher-order finite differences, and magnetic-initialization comparisons. The central test is the contrast between the identically vanishing third-order finite difference of any quadratic objective and the finite nonadditivity found in DFT energies and represented by cubic HUBO fits. We apply this comparison in both LMOF and LMTiOF, then examine how the LMTiOF difference changes with electronic initialization. In this setting, first-principles identification means detecting components outside the quadratic function space and quantifying the improvement from cubic terms on the stated domains. Magnetic-state dependence is assessed operationally through different initializations and the resulting converged solutions.

## 2 Computational methods for Quantum ESPRESSO DFT energy maps

### 2.1 Cells and composition preserving configuration spaces

Both materials were modeled in 24-atom (four-formula-unit) cells: $Li_8Mn_4O_8F_4$ for $Li_2MnO_2F$ (LMOF) and $Li_8Mn_2Ti_2O_8F_4$ for $Li_2Mn_{0.5}Ti_{0.5}O_2F$ (LMTiOF). Ideal rocksalt sites were arranged in a 3 × 2 × 2 expansion of the face-centered-cubic primitive cell. The lattice parameter was fixed at a = 4.146 Å, consistent with the calculated LMOF value reported by Sharpe et al. [22]. Using the same lattice for LMTiOF was a modeling choice, not a Ti-containing-cell optimization. The lattice vectors (Å; rows) are

$$A = \begin{pmatrix} 0 & 6.219 & 6.219 \\ 4.146 & 0 & 4.146 \\ 4.146 & 4.146 & 0 \end{pmatrix}$$

Here, 6.219 Å = 3a/2. Spin-polarized single-point SCF calculations used fixed cell vectors and atomic coordinates, without structural relaxation. Energy differences therefore compare configurations at the same composition and geometry, not formation energies or relaxed phase stabilities.

Figure 1 shows representative LMOF (01010110) and LMTiOF (001110) configurations, visualized with identical settings in VESTA 3.5.8 [23] using the fixed QE input geometries.

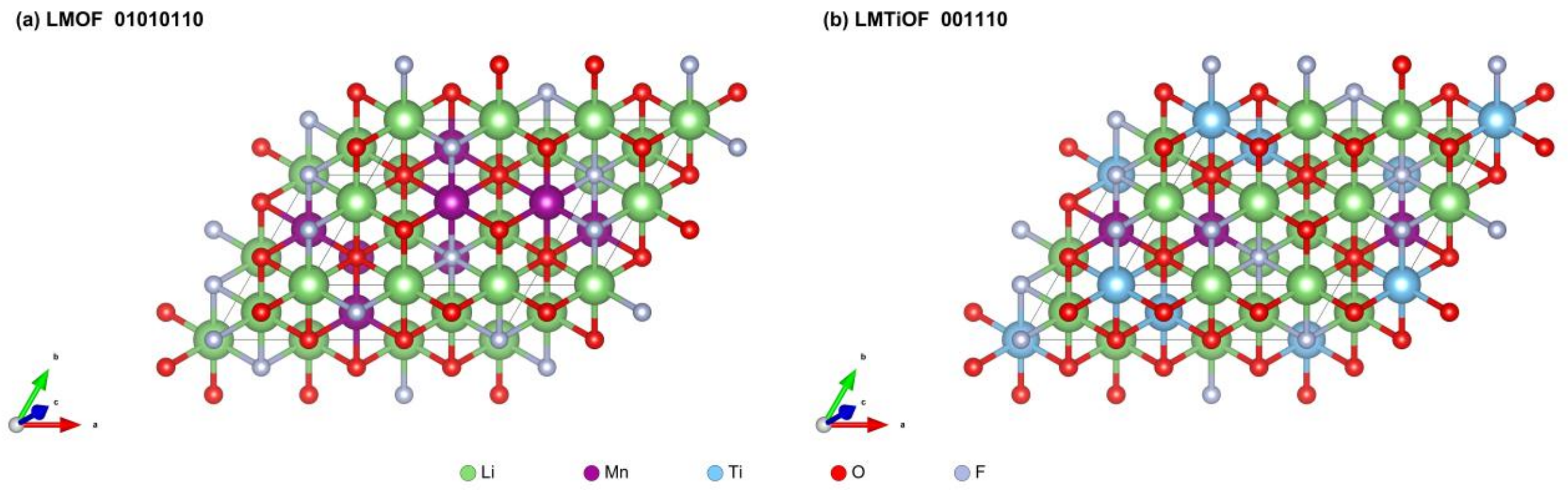

**Figure 1. Fixed-geometry 24-atom P1 models of (a) LMOF (01010110; $Li_8Mn_4O_8F_4$) and (b) LMTiOF (001110; $Li_8Mn_2Ti_2O_8F_4$), rendered from LMOF_01010110.cif and LMTiOF_001110.cif using VESTA 3.5.8 [23]. Orientation, orthographic projection, scale, and fractional-coordinate boundaries (0–1) are identical; periodic images are shown. Li, Mn, Ti, O, and F are green, purple, light blue, red, and pale lavender, respectively. Atomic positions are fixed QE input coordinates, not relaxed structures.**

For LMOF, four Li and four Mn atoms occupy eight variable cation sites; four additional Li sites and all anions remain fixed. This gives 70 allowed assignments. Eight-bit configuration labels use 1 for Li and 0 for Mn, whereas the fitting basis uses Mn-occupation indicators. Global fits are restricted to these 70 fixed-composition states. For the local $\Delta_3$ test, three composition-preserving Li–Mn exchanges at label positions 2↔3, 4↔7, and 6↔8 (1-based) generate eight configurations, from the 000 reference 00111100 to the 111 state 01001011. This eight-state exchange face is only a subset of the 70-state fitting domain.

For LMTiOF, six binary variables $x_i$ specify disjoint cation or anion exchanges relative to an all-zero reference (Table 1). All 64 combinations preserve composition and uniquely define ionic configurations.

**Table 1. Disjoint site exchanges defining the LMTiOF configuration variables.**

| Variable | Site indices (1-based) | Reference species |
|---|---|---|
| $x_1$ | 1 ↔ 5 | Mn ↔ Li |
| $x_2$ | 3 ↔ 6 | Ti ↔ Li |
| $x_3$ | 13 ↔ 15 | O ↔ F |
| $x_4$ | 2 ↔ 8 | Mn ↔ Li |
| $x_5$ | 4 ↔ 7 | Ti ↔ Li |
| $x_6$ | 14 ↔ 18 | O ↔ F |

Site indices are 1-based and follow the QE atomic-coordinate order.

The first three exchanges form an eight-state face ($x_4 = x_5 = x_6 = 0$); fixing only $x_6 = 0$ gives a 32-state, five-exchange subset of the full 64-state domain. Thus 00111 (five-variable notation) and 001110 (six-variable notation) represent the same configuration. The subsets are not independent samples, and fitted coefficients can differ between the 32- and 64-state domains. These domains exhaust only the specified exchanges, not all possible disorder arrangements.

Here, configuration denotes an ionic assignment, whereas electronic solution denotes the spin-polarized SCF state reached from a particular initialization. Different magnetic starts may therefore yield different energies for the same configuration.

Figure 2 distinguishes the full 70-state LMOF fitting domain from its eight-state $\Delta_3$ test and shows the six independent LMTiOF exchanges defining the 64-state domain.

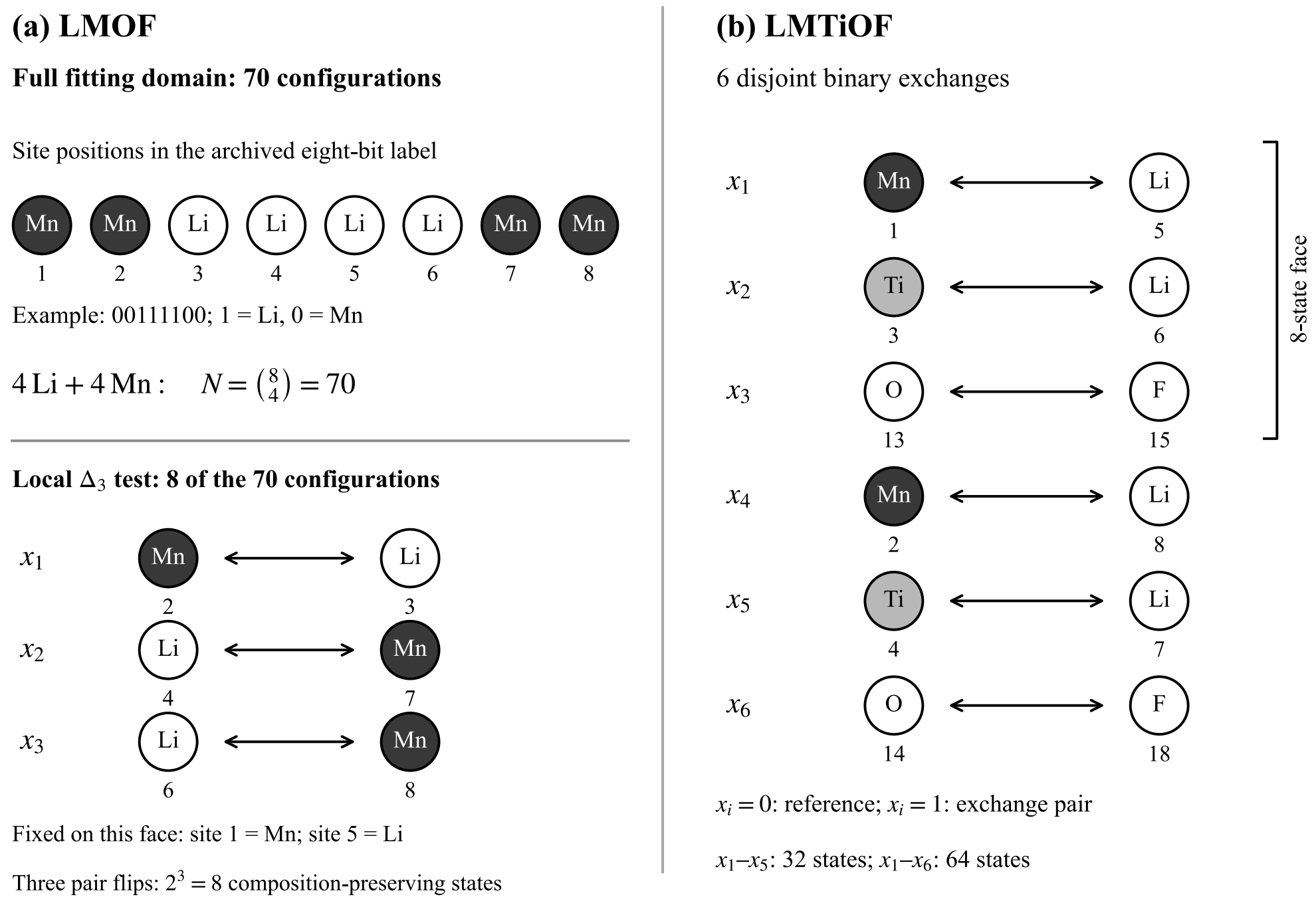

**Figure 2. Composition-preserving configuration spaces. (a) LMOF: 70 assignments of four Li and four Mn across eight variable sites (1 = Li, 0 = Mn); three specified Li–Mn exchanges (2↔3, 4↔7, and 6↔8) select eight configurations for the local $\Delta_3$ test. (b) LMTiOF: six disjoint exchange pairs (Table 1) define 8-, 32-, and 64-state nested domains. Arrows denote site reassignments, not migration paths.**

## 2.2 Electronic structure calculations

Total energies were calculated using Quantum ESPRESSO PWSCF v6.7MaX [24,25], the PBE exchange–correlation functional [26], collinear spin polarization, and PSLibrary projector-augmented-wave pseudopotentials [27]. The wavefunction and charge-density cutoffs were 110 and 1080 Ry. The automatic k-point mesh was unshifted and contained 2 × 3 × 3 points before symmetry treatment; symmetry reduction was disabled. Marzari–Vanderbilt smearing [28] was applied with a width of 0.01 Ry. The energies analyzed are the final converged total-energy records at this smearing, without a zero-smearing extrapolation.

The primary workstation used an ASUS WS X299 SAGE platform with an Intel Core i9-10900X processor (10 cores, 20 threads), 256 GB RAM, and two NVIDIA GeForce RTX 3090 GPUs (24 GB memory each), running Ubuntu Linux. The reproducibility of the LMOF calculations was assessed using SCF energies independently recalculated on a separate supercomputing system. An additional reproducibility check was performed for the eight LMTiOF three-exchange configurations on a separate desktop PC running Quantum ESPRESSO 6.7MaX under Ubuntu 22.04 in Windows Subsystem for Linux (WSL).

The simplified rotationally invariant Hubbard correction was applied to Mn with U = 3.9 eV, lda_plus_u_kind = 0, and U_projection_type = 'ortho-atomic'. No explicit U was assigned to Ti. Wang, Maxisch, and Ceder [29] examined GGA+U corrections for transition-metal oxide oxidation energies. Jain et al. [30] subsequently developed a mixed GGA/GGA+U framework for formation enthalpies, from which the numerical Mn U = 3.9 eV used here was adopted. The linear-response approach of Cococcioni and de Gironcoli [31] provides a framework for independently determining the Hubbard U parameter.

The primary LMTiOF five- and six-exchange calculations used an SCF convergence threshold of $10^{-6}$ Ry; selected states were checked at $10^{-8}$ Ry. The accepted primary LMOF calculations used $10^{-6}$ Ry for 65 configurations and $10^{-5}$ Ry for five configurations. Davidson diagonalization was used, with the conjugate-gradient method used when needed for convergence. Accepted outputs contained an explicit

SCF convergence statement, a final total energy, and successful termination. Energy differences and fitting errors are reported per 24-atom cell.

### 2.3 Magnetic initialization and the definition of an energy label

The primary LMTiOF calculations used Mn starting polarizations of the same sign, +0.55 unless otherwise stated. An additional +0.30 starting value was tested in the initial three-exchange calculation. These are two ferromagnetic-like (FM-like) initialization seeds of different magnitude, not independent magnetic configurations. The later antiferromagnetic-like (AFM-like) calculations used oppositely signed Mn starting polarizations, +0.55 and −0.55. The numerical starting polarizations are input parameters, not prescribed final magnetic moments in μB.

FM-like and AFM-like therefore name initialization classes. They do not assert that every converged state has the corresponding long-range magnetic order. Different occupation assignments can alter the final moments, electronic occupations, and basin reached from nominally identical starting values. The total and absolute magnetizations provide useful indicators, but neither quantity reconstructs the complete pattern of local Mn moments or orbital occupations. Local spin analysis would be needed to identify a microscopic magnetic arrangement.

We distinguish three energy-label sets wherever paired results are available: the FM-initialized label, the AFM-initialized label, and the lower of the two converged energies for each configuration. The last set is a two-start lower envelope. It improves the available variational selection within the tested starts but is not a global minimization over all spin and orbital solutions. A missing AFM calculation is not replaced by a predicted AFM energy in the paired analysis.

Accordingly, each energy record has two parts of its definition: the ionic bitstring and the electronic preparation used to obtain a converged label. Our magnetic comparison keeps the ionic assignment fixed and changes the starting Mn polarization. A resulting energy difference can diagnose sensitivity to the selected self-consistent solution, but the starting label alone cannot distinguish a change in local spin alignment from coupled changes in charge or orbital occupation. We therefore use magnetic initialization as the operational variable and reserve microscopic magnetic assignments for future state-resolved analysis.

### 2.4 QUBO and cubic HUBO energy representations

Ordinary least squares fits every symmetry-unrestricted monomial up to the selected degree. With distinct indices in each product, the quadratic and cubic forms are

$$E_2(x) = c_0 + \sum_i h_i x_i + \sum_{i<j} J_{ij} x_i x_j \tag{1}$$

$$E_3(x) = E_2(x) + \sum_{i<j<k} K_{ijk} x_i x_j x_k \tag{2}$$

For each degree d, the design matrix contains one row per admissible configuration and columns for the intercept and all distinct-variable monomials through degree d. Its coefficient vector collects the corresponding $c_0$, $h_i$, $J_{ij}$, and, for d = 3, $K_{ijk}$ terms. The fitted vector is the minimum-norm ordinary-least-squares solution:

$$\hat{\theta}_d = \Phi_d^{+} \mathbf{E}^{\mathrm{DFT}} \in \arg\min_{\theta} \left\| \Phi_d \theta - \mathbf{E}^{\mathrm{DFT}} \right\|_2^2, \quad d = 2, 3 \tag{2a}$$

Here the response vector contains the DFT energy labels and the superscript + denotes the Moore–Penrose pseudoinverse of the design matrix. This notation specifies the minimum-norm coefficient gauge when the site basis is rank deficient; it does not impose a regularization penalty. QUBO and cubic HUBO are fitted independently to the same labels and domain, so the cubic fit can also change the lower-order coefficients. Their full-fit errors are evaluated as

$$\mathrm{RMSE}_d = \sqrt{\frac{1}{N}\sum_{n=1}^{N}\left[E_d\left(\mathbf{x}^{(n)}\right) - E_n^{\mathrm{DFT}}\right]^2} \tag{2b}$$

where n indexes the N configurations and the model energy is evaluated at the corresponding binary assignment. The energy units are the same as those of the DFT labels.

The coefficients depend on the domain, variable convention, and selected DFT labels. The fit includes an intercept, no distance cutoff, and no regularization. An arbitrary common energy shift is absorbed by the intercept and leaves residuals and rankings unchanged. $E_2$ has QUBO form, while $E_3$ is the cubic truncation of a HUBO representation [21]. The names describe the polynomial objectives; all fits and minimum comparisons retain the admissible composition-preserving domains, including the restricted LMOF sector. On any three independently switchable, composition-preserving exchange variables, $E_2$ has $\Delta_3 = 0$ identically; small nonzero values calculated from rounded fitted energies are numerical artifacts, not a physical QUBO third-order interaction. $E_3$ can give nonzero $\Delta_3$ because it includes a cubic monomial in these exchange variables. Thus the QUBO-to-HUBO comparison concerns the order required to represent the DFT energies, with no annealing or optimizer benchmark performed.

For three, five, and six exchange variables, the $E_2$/$E_3$ design ranks are 7/8, 16/26, and 22/42. The three-variable cubic fit is saturated, with as many independent parameters as energy labels. The larger cubic fits are not saturated. In the fixed-composition LMOF sector, the corresponding ranks are 28 and 56. Site monomials are redundant on this restricted domain, so a minimum-norm coefficient vector is used. Equivalent coefficient gauges give the same fitted values within the enumerated sector; the individual coefficients should not be treated as uniquely identified unconstrained-site interactions. The 67 available SCF energies from the supercomputer are sufficient to estimate the rank-56 cubic model on the restricted fixed-composition sector, but the auxiliary fit has fewer observations than the primary 70-state fit.

The use of unrestricted monomials makes this an explicit finite-domain test of polynomial order. It differs from a transferable cluster expansion constructed from symmetry-equivalent spatial clusters and trained over many cells. Sparse and Bayesian compressive-sensing approaches offer established ways to control larger cluster bases [32,33]. The available maps instead permit direct comparison of the residual structure after projection onto two specified nested polynomial spaces.

## 2.5 Error measures and leave-one-out cross-validation

Full-fit root-mean-square error (RMSE) measures the difference between fitted and DFT energies when all N configurations are used for fitting. Leave-one-out cross-validation (LOOCV) instead omits one configuration and its DFT energy, refits the same polynomial model to the remaining N - 1 configurations, and predicts the omitted energy [34,35]. Repeating this procedure for every configuration gives N held-out residuals; their root-mean-square value is the LOOCV RMSE. For example, each LMOF fold fits 69 of the 70 configurations and predicts the remaining one. LOO abbreviates leave-one-out and is used as a subscript in Eq. (3); it denotes the same omission procedure, not a separate validation method. With the model-minus-DFT sign convention, $r_i$ is the full-fit residual and $r_{i,\mathrm{LOO}}$ is the residual predicted by the model fitted without configuration i. The leave-one-out residual follows the standard prediction sum-of-squares (PRESS) identity for ordinary least squares [35,36]:

$$r_{i,\mathrm{LOO}} = \frac{r_i}{1-h_i},$$

$$\mathrm{RMSE}_{\mathrm{LOO}} = \mathrm{RMSE}_{\mathrm{fit}}\,\frac{N}{N-p} \qquad \left(h_i = \frac{p}{N}\ \text{for every configuration}\right). \tag{3}$$

In Eq. (3), $h_i$ is the diagonal leverage of the least-squares projection matrix, not the linear energy coefficient bearing the same symbol in Eq. (1). The residual identity applies when omission preserves the design rank and $h_i < 1$. On a complete cube containing all monomials through a fixed degree, symmetry makes $h_i$ constant and equal to p/N, where p is the design rank and N the number of configurations. The fixed-composition LMOF sector has the analogous permutation symmetry. Substituting $h_i = p/N$ into the PRESS residual identity and taking the root mean square gives the second line of Eq. (3). Thus the complete-map LOOCV scores in Table 2 are deterministic rescalings of full-fit RMSE, up to rounding. They measure omitted-configuration prediction within the same energy map, but add no independent numerical test beyond those residuals and the design geometry.

For the incomplete 31-state paired set, leverage is nonuniform and the single common rescaling is inapplicable. The retained states still belong to the same small, adaptively investigated configuration family. Their LOOCV is a within-family diagnostic rather than validation on a new cell, new composition, relaxed geometry, or independently selected magnetic surface. Cubic LOOCV is not reported for the saturated eight-state cube because omitting a point leaves its full cubic model unidentifiable.

## 2.6 Direct nonadditivity and a fixed charge reference

We use the established third-order mixed finite difference, denoted here by $\Delta_3$, to quantify nonadditivity among three independently switchable site-exchange variables. This contrast is the three-variable form of the Möbius coefficient of a pseudo-Boolean function [21,37,38]. With the remaining variables fixed, it is evaluated from the same eight vertices using DFT energies or fitted-model predictions:

$$\begin{aligned}\Delta_3 = E_{111} - E_{110} - E_{101} - E_{011} \\ + E_{100} + E_{010} + E_{001} - E_{000}.\end{aligned} \tag{4}$$

The same contrast can be written as the change in the interaction between exchanges $x_1$ and $x_2$ when exchange $x_3$ is activated:

$$\Delta_3 = \underbrace{(E_{111} - E_{101} - E_{011} + E_{001})}_{\text{pair interaction at } x_3=1} - \underbrace{(E_{110} - E_{100} - E_{010} + E_{000})}_{\text{pair interaction at } x_3=0}. \tag{4a}$$

On this three-variable face, $\Delta_3$ equals the coefficient $K^{\text{face}}_{ijk}$ of $x_i x_j x_k$ in the unique exact multilinear interpolant.† Constant, linear, and pair terms cancel, so every quadratic objective in these exchange variables has $\Delta_3 = 0$. The direct DFT contrast requires no regression. Applying the contrast to cubic-model predictions returns that model's face coefficient, which need not equal the direct DFT value when the fit uses a larger map. For a higher-degree polynomial, fixed nonzero background variables can contribute higher-order terms to this effective coefficient. For LMOF, it belongs to the local exchange coordinates, not to a uniquely selected coefficient in the redundant eight-site occupation basis. It does not by itself identify a microscopic three-atom interaction or the lowest-energy configuration.

The same eight-state construction is established in thermodynamic mutant-cycle analysis, where a third-order coupling measures how a pair interaction changes in the presence of a third perturbation [39,40]. Here the perturbations are composition-preserving site exchanges and the response is the DFT configurational energy. Our contribution is its application to the specified oxyfluoride energy maps and

† For binary exchange coordinates $x_i, x_j, x_k \in \{0,1\}$ at fixed background, the exact eight-vertex interpolant has the form below. Define $\Delta_a f = f(x_a = 1) - f(x_a = 0)$. Successive differences give

$$E_{\text{face}} = c + h_i x_i + h_j x_j + h_k x_k + J_{ij} x_i x_j + J_{ik} x_i x_k + J_{jk} x_j x_k + K^{\text{face}}_{ijk} x_i x_j x_k,$$

$$\Delta_i \Delta_j E_{\text{face}} = J_{ij} + K^{\text{face}}_{ijk} x_k, \qquad \Delta_k \Delta_i \Delta_j E_{\text{face}} = K^{\text{face}}_{ijk}.$$

Expanding the three differences gives the eight signed energies in Eq. (4). The superscript "face" identifies this restricted-coordinate coefficient; equality to a fitted coefficient is exact for the fitted energy function, not necessarily for its DFT training labels.

electronic-initialization protocols. Normalization depends on the variable convention: with $s_a = 2x_a - 1$, the conventional three-factor effect is $\Delta_3/4$ and the coefficient of $s_i s_j s_k$ is $\Delta_3/8$ [41,42].

This diagnostic has a useful electrostatic reference. At fixed coordinates, a fixed-charge assignment for disjoint swaps can be written as $q(x) = q_0 + Ax$. A conventional periodic pairwise electrostatic energy has a quadratic form in q, with geometry-dependent coefficients. Substituting the affine charge relation produces a polynomial of degree at most two in x, including self and background terms under a consistent neutral-cell convention. Its $\Delta_3$ is therefore exactly zero. This analytical reference establishes the quadratic baseline for the finite-difference test.

The finite difference is also sensitive to the energy-label definition. If the eight energies have errors $\delta E$ relative to a chosen target surface, the induced change in $\Delta_3$ is their signed sum. A common offset cancels. A bound of $\varepsilon$ on every individual error implies a bound of $8\varepsilon$ on the finite difference. A standard deviation of $\sqrt{8}\sigma$ follows only for equal-variance, uncorrelated random errors. Electronic-basin changes have not been established to satisfy those assumptions, so observed branch shifts are not converted into statistical error bars.

# 3 Results

## 3.1 LMOF and LMTiOF configurational energy maps

The primary calculations provide energies for all 70 LMOF assignments and all 64 LMTiOF exchange configurations under their reported thresholds. The 32-state LMTiOF results are the $x_6 = 0$ restriction of the latter family. Table 2 summarizes the full-map representation errors, cross-validation diagnostics, and selected minima. Cubic terms reduce full-fit RMSE in both materials: from 90.18 to 27.89 meV per cell for LMOF, from 112.18 to 21.20 for the five-exchange LMTiOF face, and from 118.48 to 33.21 for the six-exchange map. We additionally compare direct DFT, QUBO, and cubic HUBO third differences on an eight-configuration face of each material. The LMOF comparison includes an independent supercomputer SCF reproducibility dataset, whose 67 records are not included in Table 2's primary-map metrics.

**Table 2. Representation errors and cross-validation for the primary DFT maps.**

| Primary map | N | Rank | | Fit RMSE | | LOOCV | | Minimum |
|---|---|---|---|---|---|---|---|---|
| | | $E_2$ | $E_3$ | $E_2$ | $E_3$ | $E_2$ | $E_3$ | |
| LMOF fixed-composition | 70 | 28 | 56 | 90.18 | 27.89 | 150.29 | 139.43 | 01010110 |
| LMTiOF 5-swap ($x_6$=0) | 32 | 16 | 26 | 112.18 | 21.20 | 224.35 | 113.07 | 00111 |
| LMTiOF 6-swap | 64 | 22 | 42 | 118.48 | 33.21 | 180.54 | 96.62 | 001110 |

Errors are in meV per 24-atom cell. $E_2$ and $E_3$ denote quadratic QUBO and cubic HUBO, respectively; rank is the design-matrix rank. LOOCV is leave-one-out cross-validation. The five-exchange face ($x_6 = 0$) is nested in the six-exchange map. The last column gives the common DFT and fitted-model minimum using the configuration labels defined in Section 2.1.

For these symmetric complete domains, $\mathrm{RMSE_{LOO}} = \mathrm{RMSE_{fit}}\,\frac{N}{N-p}$, with p the design rank. Values may differ slightly if reconstructed from rounded entries. Minimum labels follow the respective configuration-label conventions defined in Section 2.1.

Both model orders select 01010110 for LMOF and 00111 or 001110 in the five- or six-variable LMTiOF notation. The LMTiOF minima are the same ionic arrangement. Improving the average energy representation therefore does not change the selected lowest configuration within any primary map.

The comparisons are most informative within each domain. LMOF varies only a constrained selection of cation occupations, while LMTiOF varies disjoint cation and anion exchanges with a different number of Mn atoms. Their absolute energy offsets are not comparable, and their RMSE values do not isolate the chemical effect of Ti substitution. Their common feature is the presence of energy structure that is represented more accurately when the polynomial space extends beyond quadratic degree.

Table 2 also distinguishes two aspects of completeness. The primary FM-like LMTiOF map covers every ionic configuration in the six-exchange cube, whereas the additional magnetic starts cover only a subset. Thus a discrete ionic minimum can be identified exactly within the primary table while its electronic energy may still be lowered by another converged start.

For the 67 common LMOF configurations, the supercomputer-minus-workstation total-energy differences have an RMSE of 2.529 meV per cell and a maximum absolute difference of 12.184 meV per cell. Removing the common median offset gives an RMSE of 2.412 meV per cell. Both recorded SCF maps select 01010110 as their lowest-energy configuration, and their direct eight-vertex $\Delta_3$ values agree closely (Table 3). These comparisons support reproduction of the sampled energy pattern and local nonadditivity; the incomplete coverage and stricter reproduction threshold preclude describing this as a fully protocol-matched 70-state repeat. Over these 67 states, the QUBO/cubic full-fit RMSE obtained from the supercomputer energies is 86.226/27.627 meV per cell and the LOOCV RMSE is 149.352/166.909 meV per cell; restricting the primary map to the same 67 labels gives 86.223/27.762 and 149.345/167.630. The cubic representation improvement is therefore reproduced, although cubic LOOCV is higher on this incomplete set, for which scores are obtained by explicit leave-one-out refitting.

Figure 3 compares the complete primary configuration-energy maps with both polynomial fits. Configurations are ranked independently by DFT energy within each material, and all three curves use that material's minimum DFT energy as their common reference. To resolve differences hidden by the overall energy range, the lower panels show the configuration-resolved energy residual, defined as

$$\varepsilon_{d,n} = E_d\left(\mathbf{x}^{(n)}\right) - E_{\mathrm{DFT}}\left(\mathbf{x}^{(n)}\right), \qquad d \in \{2,3\} \tag{5}$$

Positive and negative residuals indicate overestimation and underestimation, respectively. The smaller cubic residuals show improved representation of the fitted maps. LMTiOF here denotes the complete 64-state FM-initialized map, without mixing in the 32-state subset, the independent local check, or alternative magnetic labels.

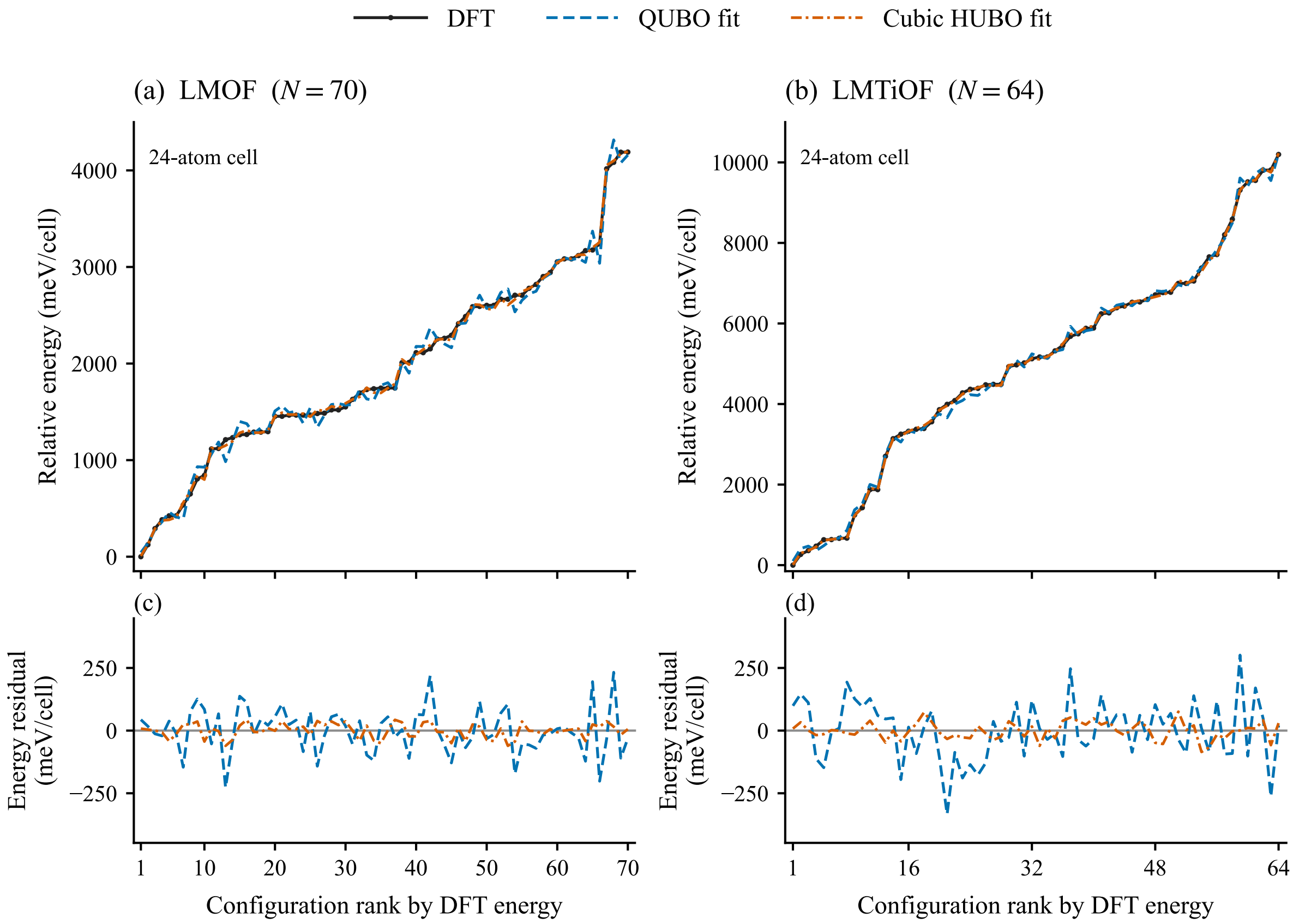

**Figure 3. Complete configuration-energy maps and fit residuals for LMOF and LMTiOF. (a,b) DFT energies (black), quadratic QUBO fits (blue dashed lines), and cubic HUBO fits (orange dash-dotted lines) for all 70 LMOF configurations and all 64 LMTiOF configurations of the primary FM-initialized map, respectively. Configurations are sorted by DFT energy within each material; each material's minimum DFT energy provides the common reference for all three curves. (c,d) Energy residuals, $\varepsilon = E_{\mathrm{model}} - E_{\mathrm{DFT}}$, where positive and negative values indicate overestimation and underestimation, respectively. Energies are per 24-atom cell. Upper-panel ranges differ; the residual panels share one range. Connecting lines guide the eye through discrete configurations and do not represent a transition path. The curves show fits to the complete primary datasets.**

## 3.2 Direct third-order nonadditivity: QUBO, HUBO, and DFT

For LMOF, the composition-preserving three-exchange face defined in Section 2.1 enables an eight-energy test within the 70 allowed cation assignments. Table 3 compares the same signed $\Delta_3$ evaluated from the original DFT energies and from quadratic and cubic fits trained on the full available data for each dataset. QUBO yields $\Delta_3 \approx 0$ numerically, as required for a degree-two polynomial. For the primary 70-state workstation SCF dataset, the cubic HUBO and direct DFT differences are −301.02 and −277.59 meV per 24-atom cell, respectively. The independent 67-state supercomputer SCF reproduction gives −299.52 and −278.05 meV per cell, respectively. The close agreement between the two DFT calculations supports reproducibility of the observed higher-order response across computational systems; it does not make the fitted cubic model an exact reproduction of the DFT energies.

**Table 3. Third-order finite differences for the LMOF exchange face.**

| DFT energy dataset | N | QUBO $\Delta_3$ | Cubic HUBO $\Delta_3$ | Direct DFT $\Delta_3$ |
|---|---|---|---|---|
| Primary SCF (workstation) | 70 | ≈ 0 | −301.02 | −277.59 |
| Independent SCF (supercomputer) | 67 | ≈ 0 | −299.52 | −278.05 |

$\Delta_3$ is in meV per 24-atom cell and is evaluated on the same eight-state exchange face. N is the number of configurations in each fitting dataset. The supercomputer SCF dataset is an independent-machine reproducibility check. QUBO gives $\Delta_3 = 0$ analytically (≈ 0 numerically).

For LMTiOF, we apply the same signed third-order difference to the original DFT energies and to QUBO and cubic HUBO predictions for the same eight-vertex Mn/Li, Ti/Li, and O/F exchange face (Table 4). The primary FM-like 64-state cubic fit gives $\Delta_3 = -109.06$ meV per cell, compared with −311.18 meV per cell from the eight original DFT energies. The nested 32-state fit and the common 31-state magnetic-comparison fits have different cubic coefficients because the fitted domains and electronic labels differ. The direct DFT finite difference on this face remains the same when the primary FM-like labels are merely restricted to a smaller fitting set. A nonzero fitted cubic $\Delta_3$ demonstrates that HUBO can capture a portion of the DFT nonadditivity, but the remaining mismatch shows that cubic truncation is not an exact representation of these larger energy maps.

**Table 4. Third-order finite differences for the LMTiOF exchange face.**

| DFT energy dataset | N | QUBO $\Delta_3$ | Cubic HUBO $\Delta_3$ | Direct DFT $\Delta_3$ |
|---|---|---|---|---|
| Primary FM-like | 64 | ≈ 0 | −109.06 | −311.18 |
| Nested FM-like | 32 | ≈ 0 | −185.33 | −311.18 |
| Matched FM-like | 31 | ≈ 0 | −172.33 | −311.18 |
| Matched AFM-like | 31 | ≈ 0 | −141.69 | −69.45 |
| Matched lower of two starts | 31 | ≈ 0 | −196.39 | −305.92 |

$\Delta_3$ is in meV per 24-atom cell and is evaluated on the same eight-state exchange face. N is the number of configurations in the fitting domain. Here, nested refers to the 32-state FM-like subset with $x_6 = 0$, while matched refers to the same 31 configurations with both FM-like and AFM-like converged energies (excluding 111100). The matched lower-of-two-starts dataset uses the lower converged energy for each of those 31 configurations. FM-like and AFM-like denote Mn-polarization initializations, not verified magnetic phases. QUBO gives $\Delta_3 = 0$ analytically (≈ 0 numerically).

In particular, the FM-like and AFM-like 31-state fits must be compared on their common 31-configuration set (missing only configuration 111100 from the 32-state five-swap face). The direct DFT values of −311.18 and −69.45 meV per cell therefore reflect sensitivity to the converged electronic solution, whereas the fitted cubic values of −172.33 and −141.69 meV per cell reflect both that sensitivity and averaging across the larger training domain.

The three-exchange face provides a separate direct algebraic test involving Mn/Li, Ti/Li, and O/F swaps. Table 5 and Figure 4 distinguish the common-threshold magnetic comparison from earlier tighter FM-like restarts. A QUBO fit to any of these energy labels gives $\Delta_3 = 0$. If a cubic model is instead fitted only to the eight vertices on this face, it reproduces the eight DFT values and their $\Delta_3$ exactly by construction; this saturated local interpolation is distinct from the full-domain HUBO fits compared in Table 4. Under the matched $10^{-6}$ Ry protocol, the direct DFT $\Delta_3$ is −311.176 meV per cell with FM-like initialization and −69.446 meV per cell with AFM-like initialization. The lower of the two tested converged energies at each of the eight vertices gives −305.922 meV per cell.

**Table 5. Direct DFT nonadditivity under different initialization protocols.**

| Input or selection protocol | $\Delta_3$ (meV/cell) | Interpretation |
|---|---|---|
| FM-like, matched $10^{-6}$ Ry | −311.176 | Original reference map |
| AFM-like, matched $10^{-6}$ Ry | −69.446 | Alternative magnetic initialization |
| Lower of two starts, matched $10^{-6}$ Ry | −305.922 | Paired, two-seed lower envelope |
| FM-like +0.55, $10^{-8}$ Ry restart | −311.362 | Tight restart, different protocol |
| FM-like +0.30, $10^{-8}$ Ry restart | −328.118 | Tight restart, not independent spin order |

$\Delta_3$ is in meV per 24-atom cell on the LMTiOF three-exchange face. The matched and tight-restart protocols are distinct; starting polarizations are input values. QUBO gives $\Delta_3 = 0$ exactly. A saturated cubic fit to each set of eight vertices reproduces the corresponding DFT energies and $\Delta_3$ by construction.

A quadratic $E_2$ (QUBO) model predicts $\Delta_3 = 0$ for this signed combination regardless of how its coefficients are fitted; a saturated three-variable cubic $E_3$ (HUBO) model reproduces the eight-point DFT $\Delta_3$ exactly. This is direct evidence that the computed energy response to the three specified exchanges cannot be summarized by their isolated and pairwise exchange contributions alone.

The finite difference can be understood as a difference of differences. First compare the energy cost of activating $x_1$ with and without $x_2$ at fixed $x_3$; then compare that pairwise response between $x_3 = 0$ and $x_3 = 1$. $\Delta_3$ measures how the interaction between the first two exchange perturbations changes when the anion exchange is activated. This construction explains why it cancels every quadratic function exactly and why a common energy reference is immaterial.

The change between the matched FM-like and AFM-like values is 241.730 meV per cell. This is a change in the nonadditive combination itself, not simply a common offset between two energy maps. Accordingly, initialization modifies the configurational dependence of the selected solutions. The lower-envelope value remains appreciably nonzero, so taking the better of the two starts does not remove the observed nonquadratic response. Its proximity to the FM value does not imply that the FM branch is lower at every vertex; a finite difference combines vertex-specific changes with alternating signs.

The earlier $10^{-8}$ Ry restarts gave −311.362 and −328.118 meV per cell for the +0.55 and +0.30 FM-like starts. These values document sensitivity within one broad initialization class and must not be combined with the later AFM comparison as though all runs sampled the same protocol. The available data identify nonquadratic labels under multiple tested preparations, while leaving the finite difference of a fully searched electronic ground-state surface unresolved.

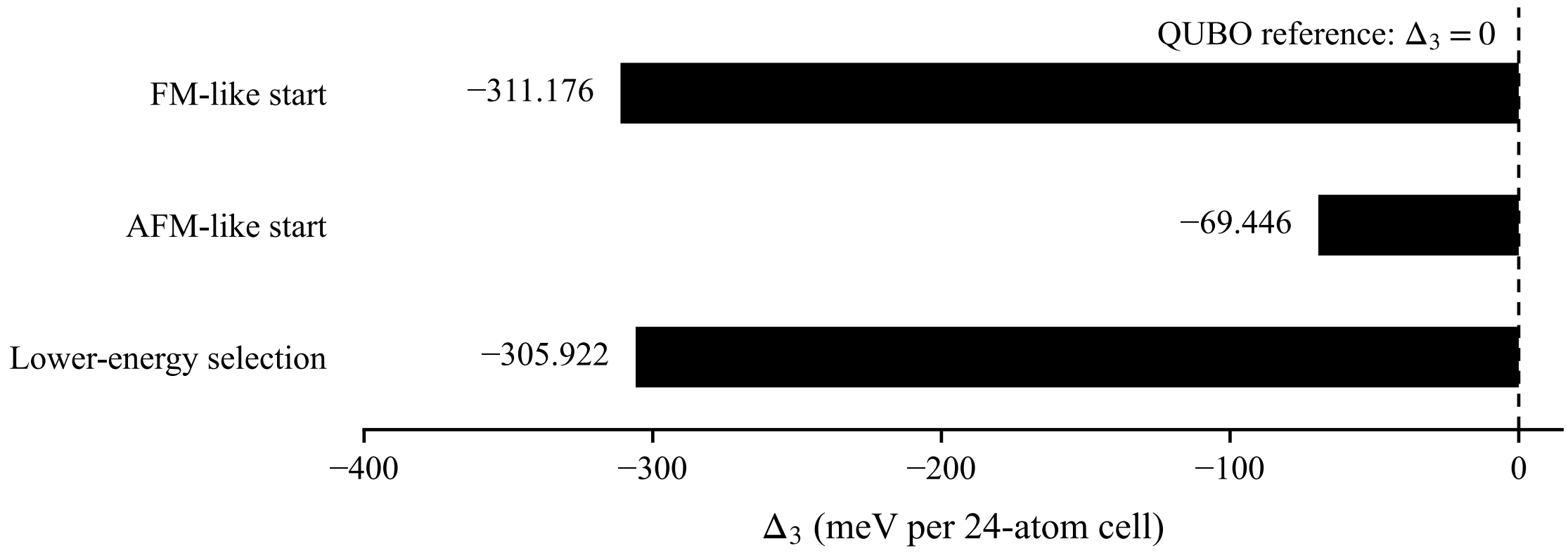

**Figure 4. Direct nonadditivity on the LMTiOF three-exchange face ($x_4 = x_5 = x_6 = 0$). Horizontal bars show $\Delta_3$ for the same eight ionic assignments under the matched $10^{-6}$ Ry protocol: −311.176 (FM-like), −69.446 (AFM-like), and −305.922 meV per 24-atom cell (lower converged energy selected at each vertex). Bars extend left from the dashed QUBO reference at $\Delta_3 = 0$; their lengths indicate the magnitude of the negative finite difference. Labels denote initialization or selection protocols, not verified magnetic phases. Tight-restart results are listed separately in Table 5; no statistical error bars are assigned.**

The QUBO-to-HUBO extension adds cubic directions to the quadratic function space, so a reduction in training residual is expected when those directions overlap the residual energy vector. The first-principles result is the magnitude of the captured configurational contribution and its persistence under the paired magnetic comparisons. Together with the model–DFT $\Delta_3$ comparisons for LMOF (Table 3) and LMTiOF (Table 4), and the initialization-resolved direct DFT tests for LMTiOF (Table 5), Figure 4 identifies energy structure that QUBO cannot represent on these domains. The nonzero cubic residuals show that the chosen HUBO truncation is still incomplete; they do not determine whether the remaining deviations mainly reflect higher polynomial degree, changes in electronic basin, or numerical protocol variation.

Figure 5 places this local diagnostic alongside the full-domain model hierarchy. The direct difference tests one specified face, while the least-squares residual averages over the entire reported domain. Together, the two analyses characterize local nonadditivity and full-domain representation error in the same DFT energy map.

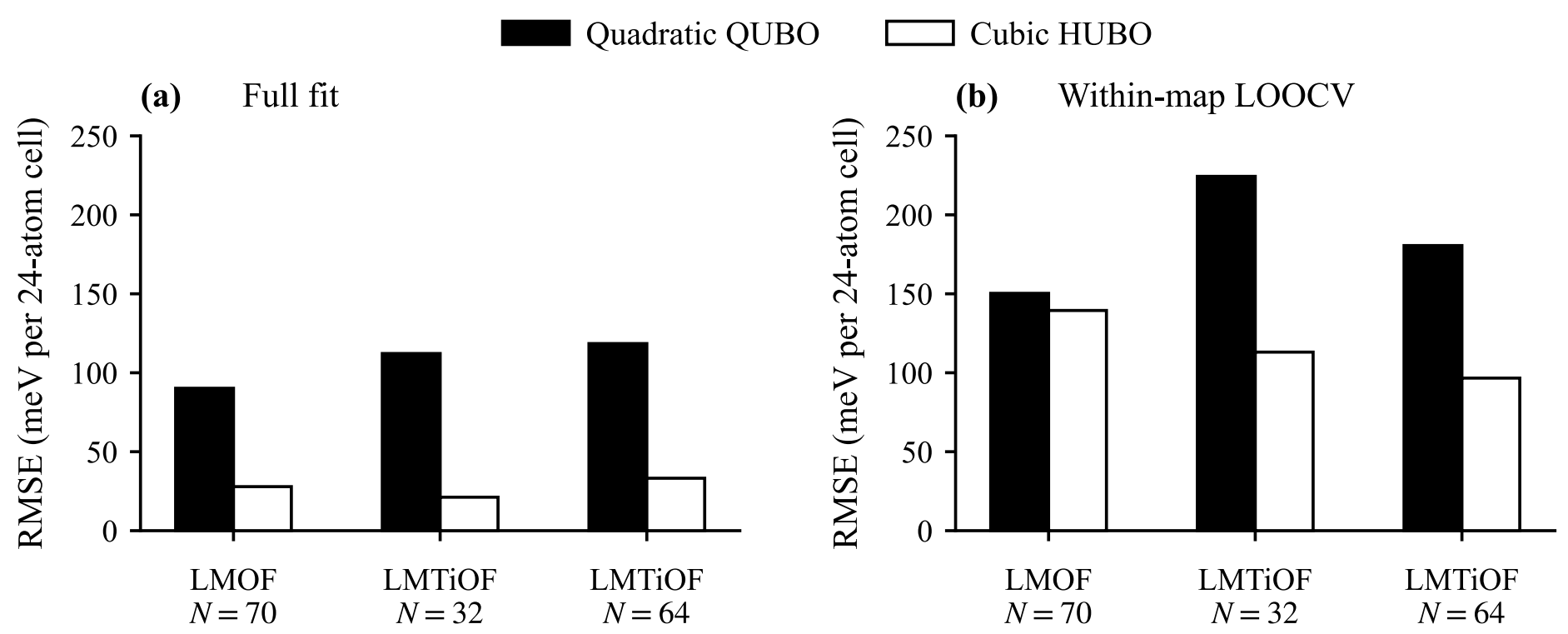

**Figure 5. Quadratic and cubic representation errors for the primary DFT maps. (a) Full-fit and (b) within-map leave-one-out cross-validation (LOOCV) RMSE for the 70-state LMOF sector and nested 32- and 64-state LMTiOF domains. In panel (b), each configuration is predicted using a fit that excludes its DFT energy. Black bars denote quadratic QUBO; white bars with black outlines denote cubic HUBO. Both panels share the same scale (meV per 24-atom cell). Cubic terms reduce but do not eliminate residuals. For these complete symmetric domains, Eq. (3) relates LOOCV algebraically to full-fit RMSE.**

## 3.3 Magnetic state dependence of configurational energies

The later AFM-like campaign produced accepted converged energies for 32 of the 64 six-exchange configurations. Within the five-exchange face, 31 of the 32 states have matched FM and AFM labels. The remaining arrangement, written 111100 in six-variable notation with $x_6 = 0$, did not reach accepted AFM convergence despite recovery attempts. The comparisons in Table 6 use exactly the same 31 configurations for each label-selection rule.

**Table 6. Paired model comparison on the 31-state LMTiOF subset.**

| 31-state label set | Fit RMSE | | LOOCV | | Minimum |
|---|---|---|---|---|---|
| | $E_2$ | $E_3$ | $E_2$ | $E_3$ | |
| Original FM-initialized labels | 102.42 | 21.16 | 217.11 | 134.12 | 00111 |
| AFM-initialized labels | 113.19 | 12.85 | 238.09 | 81.46 | 00111 |
| Lower of the two converged starts | 104.53 | 14.93 | 221.99 | 94.68 | 00111 |

Errors are in meV per 24-atom cell. All three label sets contain the same 31 ionic configurations on the five-exchange subset. $E_2$ and $E_3$ denote quadratic QUBO and cubic HUBO, respectively. LOOCV is leave-one-out cross-validation.

On this common subset, the FM-initialized fit RMSE decreases from 102.42 to 21.16 meV per cell when cubic terms are added. The corresponding AFM-initialized values are 113.19 and 12.85, and the two-start lower-envelope values are 104.53 and 14.93. Cubic terms thus retain a substantial descriptive contribution after controlling the set of ionic assignments and changing the initialization class. The effect is not confined to fitting the original FM labels.

The LOOCV values also decrease for each paired label set: from 217.11 to 134.12 meV per cell for FM-like starts, from 238.09 to 81.46 for AFM-like starts, and from 221.99 to 94.68 for the lower envelope. These values should be considered alongside the design ranks, 16 for $E_2$ and 26 for $E_3$ on only 31 states. The remaining sample count is small relative to cubic model complexity, and leverage can amplify held-out errors. A small cubic training error consequently does not imply equally small omitted-point error.

Among all 32 accepted AFM comparisons, 22 AFM-initialized energies are lower than their FM counterparts. Examples of AFM-minus-FM shifts are −165.69 meV per cell for 010101, −107.43 for 001100, and −71.30 for 101110. These examples use six-variable labels. They establish that using a single same-sign Mn start can leave materially lower converged solutions unexplored. They do not provide a probability of finding the lowest electronic state, because the configurations with accepted AFM solutions are not a statistically complete sample of possible magnetic states.

The original FM, AFM, and lower-envelope fits on the 31-state subset all select 00111 as the lowest five-variable configuration. This agreement concerns the available paired face. It does not rule out a still lower AFM solution for the unresolved state or an untested spin/orbital preparation elsewhere in the 64-state family.

The paired analysis answers a narrower question than which magnetic state is stable in the material. On the common ionic subset, the preference for cubic order persists for FM-initialized labels, AFM-initialized labels, and their two-start lower envelope. This persistence supports the descriptive need for terms beyond quadratic order under all three available labeling rules. In contrast, the large change in $\Delta_3$ shows that its numerical coefficient cannot be transferred from one rule to another without recalculation. The model hierarchy is more consistent across these tests than the magnitude of an individual higher-order descriptor.

### 3.3.1 Lower electronic solutions and magnetic coverage

A sensitivity calculation on the complete 64-state grid replaces an FM energy by an accepted lower AFM energy wherever available, retaining the FM label elsewhere. Its quadratic/cubic fit RMSE is 114.42/29.74 meV per cell, and its LOOCV RMSE is 174.35/86.50. The corresponding original-FM values are 118.48/33.21 and 180.54/96.62. This mixed-coverage result shows that the residual hierarchy persists under the presently available replacements. It is not a uniformly paired magnetic map and is not included as another independent dataset in Table 2.

An independent $10^{-8}$ Ry AFM-like calculation for configuration 001110 converged to −1633.78490954 Ry, lying 39.948 meV per cell below the existing tightly converged FM-like solution. Its total magnetization is approximately −0.02 μB per cell and its absolute magnetization is 9.77 μB per cell. The small total and appreciable absolute magnetization are compatible with cancellation of local spin densities.

The lower AFM solution is significant because the ionic arrangement is the same as the primary enumerated minimum. A tighter SCF solution can therefore remain electronically metastable relative to another converged initialization. It demonstrates that electronic-state selection is part of the energy-map construction and that complete configuration coverage should be reported separately from magnetic-start coverage.

### 3.3.2 Numerical checks and reproducibility

Selected fresh tighter-SCF calculations had previously changed LMTiOF energies by roughly 107–130 meV per cell while also changing the final total moments. Other tested states were essentially unchanged. Such behavior supports electronic-solution sensitivity rather than treating the nominal SCF threshold as a direct error bar on every energy. Tightening the stopping criterion and changing the converged basin are distinct operations, even when they occur in the same restart.

An additional independent-installation check of the eight LMTiOF three-exchange configurations was performed using PWSCF 6.7MaX under Ubuntu 22.04 in Windows Subsystem for Linux on a Windows 11 PC with an Intel Core i5-14400 processor (10 cores, 16 threads) and 16 GB RAM. Seven energies reproduced the primary preliminary results within 0.054 meV per cell, whereas configuration 101 differed by 21.153 meV per cell and its total moment changed from 8.86 to 9.16 μB, indicating sensitivity to the converged electronic state.

The finite-difference observations must consequently be interpreted at the scale of the demonstrated branch changes. SCF convergence establishes self-consistency for an accepted solution, whereas systematic convergence of relative energies and comprehensive electronic-state selection address different questions. Noninteger total moments at finite smearing further motivate spin-resolved density-of-states, band-gap, local-moment, and occupation checks. Those quantities have not been established by the energy tables, and a metallic or insulating state is not assigned from the total moment alone.

# 4 Discussion

## 4.1 Implications of higher-order configurational energetics

The DFT maps show two complementary signatures of higher-order energetics: a large residual reduction when $E_2$ is extended to $E_3$, and a directly nonzero third-order difference on a defined exchange face. The first is an average error over a finite domain; the second is a single signed combination of energies. Applied to the fitted QUBO, this combination gives zero, whereas the DFT data and the cubic fits give substantial nonzero values. Because the LMOF cubic $\Delta_3$ is close to, but not equal to, the direct DFT value, model-based and data-based diagnostics are reported separately.

Both quantities depend on the chosen exchange basis. The LMTiOF $\Delta_3$ is evaluated at $x_4 = x_5 = x_6 = 0$, and at a nonzero background it could also contain fourth- and higher-degree contributions; the LMOF diagnostic uses the three cation swaps defined in Section 2.1. Neither value is a background-independent interaction parameter. Conventional cluster expansions address spatial symmetry, cluster multiplicity, basis selection, and training-domain coverage [6–9], whereas the present representation deliberately fixes a small exchange domain to make nonadditivity explicit. The result is therefore an effective higher-order configurational contribution under a specified DFT protocol, not a transferable lattice-cluster interaction.

### 4.2 Electronic basin selection and magnetic coupling

The change in $\Delta_3$ from −311 to −69 meV per cell between FM-like and AFM-like initializations on the same 31 configurations shows that the higher-order coefficients are not purely chemical. The DFT energy is a function E(x,s) of occupations x and electronic or magnetic state s, so any occupation-only label set embeds a choice of s, either through a fixed initialization protocol or through a minimum over the tested starts. Even if each branch has simple occupation dependence, switching between branches as x changes generates higher-order terms in the effective energy. The fitted cubic coefficients therefore combine chemical response with branch selection and should be read as protocol-conditioned descriptors. The two-start envelope is also not the lowest-energy surface $E^*(x) = \min_s E(x, s)$. Because Eq. (4) contains terms of both signs, lowering the energy of one configuration does not fix the sign or magnitude of the resulting change in $\Delta_3$. A model of the fixed-geometry ground state would require a consistent electronic-minimization rule, whereas a branch-specific model would require the branch to be tracked explicitly.

A plausible physical origin is occupation-dependent Mn–Mn exchange, in which a coupling J(x) that depends on Mn placement and O/F surroundings multiplies a spin correlation $S_1 \cdot S_2$ [10,11]. The present data do not separate this mechanism from charge redistribution, hybridization changes, or alternative orbital occupations. In addition, the three-exchange face varies one Mn/Li, one Ti/Li, and one O/F exchange and was not designed to probe a specific Mn–anion–Mn superexchange path.

### 4.3 Fixed geometry and the electrostatic reference

At fixed coordinates, a fixed-charge pairwise electrostatic energy is exactly quadratic in the exchange variables (Section 2.6) and therefore gives $\Delta_3 = 0$. The nonzero DFT values thus require physics beyond fixed point charges, such as configuration-dependent charge redistribution, polarization, or electronic screening. This does not make electrostatics irrelevant; it rules out only its fixed-charge pairwise form as the source of the third-order term.

This null result holds only at fixed geometry. If atomic positions relax, the ionic coordinates themselves depend on x, and even pairwise electrostatics can generate a nonzero $\Delta_3$. Relaxation may therefore reduce, enhance, or redistribute the higher-order response, and the coupling between local configuration and distortion reported for oxyfluorides [5] suggests that this effect is material. Fixed geometry isolates the electronic contribution but retains post-exchange strain; comparing the same configurations before and after relaxation would separate the structural and electronic parts of the nonadditivity.

### 4.4 Representation accuracy and predictive use

These results concern the representation of a DFT energy objective, not its search. Both materials show a nonzero direct $\Delta_3$ that any QUBO in the original variables misses, and cubic HUBO captures part or all of it depending on the fit domain. This does not exclude an exact quadratization with auxiliary variables [43]; direct HUBO treatment and quadratized QUBO differ in resource cost [18,19], a comparison not attempted here. Likewise, the quantum-algorithm studies cited above [15–17,20] provide context rather than a performance benchmark for these materials.

The full-fit and LOOCV errors also answer different questions. For LMOF, the cubic fit greatly reduces the full-map error, but fitting 56 independent directions to 70 states produces large leverage, so the LOOCV gain is modest; this does not negate the nonquadratic component. Conversely, the larger LMTiOF LOOCV reductions do not by themselves demonstrate transferability. A predictive claim would require configurations withheld from both fitting and model selection, and a minimum-discovery claim would require DFT verification of model-selected candidates. The present scope is one fixed geometry per material, a single functional and Hubbard setup, finite smearing, and incomplete magnetic-start coverage.

## 5 Conclusion

DFT calculations for $Li_2MnO_2F$ (LMOF) and $Li_2Mn_{0.5}Ti_{0.5}O_2F$ (LMTiOF) reveal configurational energy contributions that quadratic QUBO models cannot fully represent. On the specified three-exchange domains, QUBO gives $\Delta_3 = 0$ by construction, whereas direct DFT energy differences yield nonzero values. Cubic HUBO models capture part of this missing contribution and improve the description of both energy maps. For LMOF, independent supercomputing calculations reproduce the nonadditivity found in the primary workstation calculations. For LMTiOF, changing the Mn magnetic initialization substantially changes the DFT $\Delta_3$, while comparisons using the same configurations and energy labels retain a clear improvement in cubic fit accuracy.

Together, these findings provide a first-principles route for identifying effective higher-order configurational couplings on the specified exchange domains and for constructing protocol-conditioned HUBO energy models. Extending this framework to relaxed structures, larger supercells, broader chemical compositions, and systematically sampled magnetic states could improve the description of short-range ordering and facilitate the identification of energetically favorable atomic arrangements in lithium-rich disordered-rocksalt cathodes. Furthermore, DFT-derived HUBO models provide a promising basis for developing and benchmarking classical and quantum optimization algorithms, including the quantum approximate optimization algorithm (QAOA), thereby connecting first-principles materials modeling with emerging quantum optimization approaches.

## Acknowledgement

This study was supported by the National Research Foundation of Korea (grant number: RS2026-25599908, THz-Correlated Nano-Optical and Surface-Spectroscopic Responses and Charge Dynamics of MXene Nanostructures: Effects of Electrochemical Modulation).

## References

[1] J. Lee, D. A. Kitchaev, D.-H. Kwon, et al. Reversible Mn2+/Mn4+ double redox in lithium-excess cathode materials. Nature 556, 185–190 (2018). https://doi.org/10.1038/s41586-018-0015-4

[2] K. McColl, R. A. House, G. J. Rees, A. G. Squires, S. W. Coles, P. G. Bruce, B. J. Morgan, M. S. Islam. Transition metal migration and O2 formation underpin voltage hysteresis in oxygen-redox disordered rocksalt cathodes. Nature Communications 13, 5275 (2022). https://doi.org/10.1038/s41467-022-32983-w

[3] L. Li, B. Ouyang, Z. Lun, et al. Atomic-scale probing of short-range order and its impact on electrochemical properties in cation-disordered oxide cathodes. Nature Communications 14, 7448 (2023). https://doi.org/10.1038/s41467-023-43356-2

[4] A. G. Squires, D. O. Scanlon. Understanding the limits to short-range order suppression in many-component disordered rock salt lithium-ion cathode materials. Journal of Materials Chemistry A 11, 13765–13773 (2023). https://doi.org/10.1039/D3TA02088F

[5] N. J. Szymanski, Z. Lun, J. Liu, E. C. Self, C. J. Bartel, J. Nanda, B. Ouyang, G. Ceder. Modeling Short-Range Order in Disordered Rocksalt Cathodes by Pair Distribution Function Analysis. Chemistry of Materials 35, 4922–4934 (2023). https://doi.org/10.1021/acs.chemmater.2c03827

[6] J. M. Sanchez, F. Ducastelle, D. Gratias. Generalized cluster description of multicomponent systems. Physica A: Statistical Mechanics and its Applications 128, 334–350 (1984). https://doi.org/10.1016/0378-4371(84)90096-7

[7] J. M. Sanchez. Cluster expansions and the configurational energy of alloys. Physical Review B 48, 14013–14015 (1993). https://doi.org/10.1103/PhysRevB.48.14013

[8] J. M. Sanchez. Cluster expansion and the configurational theory of alloys. Physical Review B 81, 224202 (2010). https://doi.org/10.1103/PhysRevB.81.224202

[9] L. Barroso-Luque, G. Ceder. The cluster decomposition of the configurational energy of multicomponent alloys. npj Computational Materials 10, 158 (2024). https://doi.org/10.1038/s41524-024-01338-y

[10] M. Yu. Lavrentiev, S. L. Dudarev, D. Nguyen-Manh. Magnetic cluster expansion simulations of FeCr alloys. Journal of Nuclear Materials 386–388, 22–25 (2009). https://doi.org/10.1016/j.jnucmat.2008.12.052

[11] G. E. García Ponte, S. S. Behara, E. N. Bassey, R. J. Clément, A. Van der Ven. First-Principles Statistical Mechanics Study of Magnetic Fluctuations and Order–Disorder in the Spinel $LiNi_{0.5}Mn_{1.5}O_4$ Cathode. Chemistry of Materials 37, 1835–1846 (2025). https://doi.org/10.1021/acs.chemmater.4c02772
[12] Y. Couzinié, Y. Nishiya, H. Nishi, T. Kosugi, H. Nishimori, Y.-i. Matsushita. Annealing for prediction of grand canonical crystal structures: Implementation of n-body atomic interactions. Physical Review A 109, 032416 (2024). https://doi.org/10.1103/PhysRevA.109.032416
[13] K. Ichikawa, G. Hayashi, S. Ohuchi, T. Yokoyama, K. N. Okada, K. Fujii. Optimal elemental configuration search in crystal using quantum approximate optimization algorithm. arXiv:2503.09356 (2025), preprint. https://doi.org/10.48550/arXiv.2503.09356
[14] K. Ichikawa, S. Ohuchi, T. Yokoyama, T. Saito, Y. Matsuda. Constrained Optimization of Higher-Order Cluster-Expansion Hamiltonians for Alloys Using Simulated Bifurcation. arXiv:2609.10310 (2026), preprint. https://doi.org/10.48550/arXiv.2609.10310
[15] E. Pelofske, A. Bärtschi, S. Eidenbenz. Short-depth QAOA circuits and quantum annealing on higher-order Ising models. npj Quantum Information 10, 30 (2024). https://doi.org/10.1038/s41534-024-00825-w
[16] E. Pelofske, A. Bärtschi, L. Cincio, J. Golden, S. Eidenbenz. Scaling whole-chip QAOA for higher-order Ising spin glass models on heavy-hex graphs. npj Quantum Information 10, 109 (2024). https://doi.org/10.1038/s41534-024-00906-w
[17] S. V. Romero, A.-M. Visuri, A. Gomez Cadavid, A. Simen, E. Solano, N. N. Hegade. Bias-field digitized counterdiabatic quantum algorithm for higher-order binary optimization. Communications Physics 8, 348 (2025). https://doi.org/10.1038/s42005-025-02270-3
[18] F. Koch, S. Panahiyan, R. Mukherjee, J. Doetsch, D. Jaksch. Resource-efficient quantum optimization via higher-order encoding. EPJ Quantum Technology 13, 59 (2026). https://doi.org/10.1140/epjqt/s40507-026-00526-7
[19] S. Nagies, K. T. Geier, J. Akram, D. Bantounas, M. Johanning, P. Hauke. Boosting quantum annealing performance through direct polynomial unconstrained binary optimization. Quantum Science and Technology 10, 035008 (2025). https://doi.org/10.1088/2058-9565/adcae6
[20] E. O. Kiktenko, E. V. Krendeleva, A. K. Fedorov. Applying a Grover-mixer quantum alternating-operator-ansatz algorithm to higher-order unconstrained binary optimization problems. Physical Review A 113, 052451 (2026). https://doi.org/10.1103/p2t5-x8l5
[21] E. Boros, P. L. Hammer. Pseudo-Boolean optimization. Discrete Applied Mathematics 123, 155–225 (2002). https://doi.org/10.1016/S0166-218X(01)00341-9
[22] R. Sharpe, R. A. House, M. J. Clarke, D. Förstermann, J.-J. Marie, G. Cibin, K.-J. Zhou, H. Y. Playford, P. G. Bruce, M. S. Islam. Redox Chemistry and the Role of Trapped Molecular $O_2$ in Li-Rich Disordered Rocksalt Oxyfluoride Cathodes. Journal of the American Chemical Society 142, 21799–21809 (2020). https://doi.org/10.1021/jacs.0c10270
[23] K. Momma, F. Izumi. VESTA 3 for three-dimensional visualization of crystal, volumetric and morphology data. Journal of Applied Crystallography 44, 1272–1276 (2011). https://doi.org/10.1107/S0021889811038970
[24] P. Giannozzi, S. Baroni, N. Bonini, et al. QUANTUM ESPRESSO: a modular and open-source software project for quantum simulations of materials. Journal of Physics: Condensed Matter 21, 395502 (2009). https://doi.org/10.1088/0953-8984/21/39/395502
[25] P. Giannozzi, O. Andreussi, T. Brumme, et al. Advanced capabilities for materials modelling with Quantum ESPRESSO. Journal of Physics: Condensed Matter 29, 465901 (2017). https://doi.org/10.1088/1361-648X/aa8f79
[26] J. P. Perdew, K. Burke, M. Ernzerhof. Generalized Gradient Approximation Made Simple. Physical Review Letters 77, 3865–3868 (1996). https://doi.org/10.1103/PhysRevLett.77.3865
[27] A. Dal Corso. Pseudopotentials periodic table: From H to Pu. Computational Materials Science 95, 337–350 (2014). https://doi.org/10.1016/j.commatsci.2014.07.043
[28] N. Marzari, D. Vanderbilt, A. De Vita, M. C. Payne. Thermal Contraction and Disordering of the Al(110) Surface. Physical Review Letters 82, 3296–3299 (1999). https://doi.org/10.1103/PhysRevLett.82.3296
[29] L. Wang, T. Maxisch, G. Ceder. Oxidation energies of transition metal oxides within the GGA+U framework. Physical Review B 73, 195107 (2006). https://doi.org/10.1103/PhysRevB.73.195107
[30] A. Jain, G. Hautier, S. P. Ong, C. J. Moore, C. C. Fischer, K. A. Persson, G. Ceder. Formation enthalpies by mixing GGA and GGA+U calculations. Physical Review B 84, 045115 (2011). https://doi.org/10.1103/PhysRevB.84.045115
[31] M. Cococcioni, S. de Gironcoli. Linear response approach to the calculation of the effective interaction parameters in the LDA+U method. Physical Review B 71, 035105 (2005). https://doi.org/10.1103/PhysRevB.71.035105
[32] L. J. Nelson, G. L. W. Hart, F. Zhou, V. Ozoliņš. Compressive sensing as a paradigm for building physics models. Physical Review B 87, 035125 (2013). https://doi.org/10.1103/PhysRevB.87.035125
[33] L. J. Nelson, V. Ozoliņš, C. S. Reese, F. Zhou, G. L. W. Hart. Cluster expansion made easy with Bayesian compressive sensing. Physical Review B 88, 155105 (2013). https://doi.org/10.1103/PhysRevB.88.155105
[34] M. Stone. Cross-Validatory Choice and Assessment of Statistical Predictions. Journal of the Royal Statistical Society: Series B (Methodological) 36, 111–133 (1974). https://doi.org/10.1111/j.2517-6161.1974.tb00994.x
[35] D. M. Allen. The Relationship Between Variable Selection and Data Augmentation and a Method for Prediction. Technometrics 16, 125–127 (1974). https://doi.org/10.1080/00401706.1974.10489157
[36] G. H. Golub, M. Heath, G. Wahba. Generalized Cross-Validation as a Method for Choosing a Good Ridge Parameter. Technometrics 21, 215–223 (1979). https://doi.org/10.1080/00401706.1979.10489751
[37] M. Grabisch, J.-L. Marichal, M. Roubens. Equivalent Representations of Set Functions. Mathematics of Operations Research 25, 157–178 (2000). https://doi.org/10.1287/moor.25.2.157.12225

[38] J.-L. Marichal, P. Mathonet. Influence and interaction indexes for pseudo-Boolean functions: A unified least squares approach. Discrete Applied Mathematics 179, 13–27 (2014). https://doi.org/10.1016/j.dam.2014.05.039
[39] F. J. Poelwijk, V. Krishna, R. Ranganathan. The Context-Dependence of Mutations: A Linkage of Formalisms. PLOS Computational Biology 12, e1004771 (2016). https://doi.org/10.1371/journal.pcbi.1004771
[40] A. Horovitz, A. R. Fersht. Strategy for analysing the co-operativity of intramolecular interactions in peptides and proteins. Journal of Molecular Biology 214, 613–617 (1990). https://doi.org/10.1016/0022-2836(90)90275-Q
[41] NIST/SEMATECH. e-Handbook of Statistical Methods, Section 5.3.3.3.2: Full factorial example. https://www.itl.nist.gov/div898/handbook/pri/section3/pri3332.htm (accessed 25 October 2026).
[42] NIST/SEMATECH. e-Handbook of Statistical Methods, Section 5.5.9.7: |Effects| plot. https://www.itl.nist.gov/div898/handbook/pri/section5/pri597.htm (accessed 25 October 2026).
[43] E. Boros, A. Gruber. On Quadratization of Pseudo-Boolean Functions. arXiv:1404.6538 (2014). https://arxiv.org/abs/1404.6538